\documentclass[preprint,aps,superscriptaddress]{revtex4}
\usepackage{graphicx}
\usepackage{dcolumn}
\usepackage{bm}
\usepackage{times}

\begin{document}

\title{Photonic crystal polarizers and polarizing beam splitters}
\author{D. R. Solli}
\author{C. F. McCormick}
\author{R. Y. Chiao}
\affiliation{Department of Physics, University of California, Berkeley, CA 94720-7300}
\author{J. M. Hickmann}
\affiliation{Department of Physics, University of California, Berkeley, CA 94720-7300}
\affiliation{Departamento de F\'{\i}sica, Universidade Federal de Alagoas, Cidade
Universit\'{a}ria, 57072-970, Macei\'{o}, AL, Brazil}

\begin{abstract}
We have experimentally demonstrated polarizers and polarizing beam splitters based on microwave-scale two-dimensional photonic crystals. Using polarized microwaves within certain frequency bands, we have observed a squared-sinusoid (Malus) transmission law when using the photonic crystal as a polarizer. The
photonic crystal also functions as a polarizing beamsplitter; in this configuration it can be engineered to split incident polarizations in either
order, making it more versatile than conventional, Brewster-angle
beamsplitters.
\end{abstract}

\maketitle

Photonic crystals are periodic dielectric structures that possess a unique
band structure for electromagnetic (EM) waves. Their periodic boundary
conditions produce multiple Bragg reflections, preventing the propagation of
radiation within certain frequency and wavevector bands. In general, the
specific characteristics of the band structure of a photonic crystal depend
strongly on its geometry and material composition. A diverse set of
applications has been proposed for photonic crystals\cite{review-reference},
including suppression of spontaneous emission \cite{Yablonovitch1987} and
optical waveguides \cite{Cregan1999,Chow2000}.

Although scalar approximations are usually adequate for electronic band
structure calculations in solid-state materials, the vector nature of EM
radiation must be taken into account when physically modelling photonic
crystals. In particular, the transmission properties of a photonic crystal
depend on its orientation relative to the polarization of the incident
fields. Polarization-dependent band gap characteristics have been
experimentally observed for propagation in the plane of periodicity in both
slab and bulk two-dimensional (2D) photonic crystals \cite%
{Chow2000,Foteinopoulou2001,Hickmann2002}. While both of these structures
possess periodicity in two dimensions, slabs are thin in their extruded
(nonperiodic) dimension compared with the lattice constant, whereas bulk
crystals are much larger in their extruded dimension than the lattice
spacing. In bulk crystals, the band gaps for both incident polarizations are
well-formed, although the center frequency, depth, width, and shape depend
on the polarization \cite{Hickmann2002}. Slab crystals have a well-defined
band gap for only one polarization and show only a small, incomplete band
gap for the other \cite{Joannopoulos1999}.

The polarization properties associated with the band structure of a 2D
crystal can be qualitatively understood considering the anisotropy of the
lattice. In particular, each polarization experiences different boundary
conditions depending on whether it is parallel or perpendicular to the plane
of symmetry. Thus, the Fresnel coefficients are different for the two
possible polarizations at each dielectric interface, even at normal
incidence. These birefrigent properties have been recently exploited in the
construction of a compact new kind of photonic crystal waveplate capable of
controlling the polarization of light in transparent spectral regions \cite%
{Solli2002b}.

In this work, we report a proof of principle experiment demonstrating a
polarizer and a polarizing beam splitter (PBS) based on a bulk 2D photonic
crystal, using microwave radiation. We show that a 2D photonic crystal can
act as a polarizer if there is sufficient separation between the band gaps for
different light polarizations. For the frequency ranges in
which the band gaps do not overlap, one polarization is transmitted while
the other is reflected. The quality of this type of polarizer depends on the
transmission contrast and absolute transmission in the frequency region in
question, and its useful bandwidth is determined by the frequency width of
the non-overlap region. Photonic crystal polarization splitters have been suggested 
before \cite{Ohtera1999}; however, emphasis was placed on the fabrication of a 
specific photonic crystal structure and no new direct experimental verification 
of the operation of the polarizer was presented. In addition, the novelty of 
a Brewster-independent polarizing beamsplitter was not discussed. 
It is also important to note that the results presented in our work were 
obtained with significantly lower index contrast in the dielectric materials.

We constructed the photonic crystals used in this experiment with a method
described elsewhere \cite{Hickmann2002}. In short, hollow acrylic pipes
(refraction index $1.61$ \cite{Gray1972}) stacked in a triangular array form
a bulk hexagonal lattice whose air-filling fraction (AFF) and lattice
spacing are determined by the inner and outer diameters of the pipes. The
crystals used in this experiment had a lattice spacing of 1/2 inch, with an
AFF of 0.60. We tested crystals ranging from two to twenty layers. The scale
of these crystals allows us to build them with extremely high precision and a
large number of layers.

\begin{figure}[h]
\centerline{\includegraphics{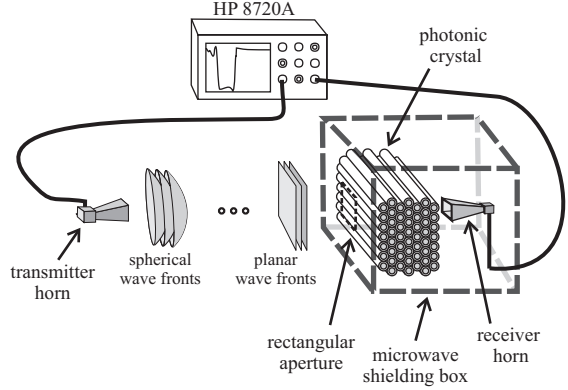}}
\caption{Experimental setup for transmission measurements.}
\label{setup}
\end{figure}

We measured the transmission through the crystals using a method that has
also been previously described \cite{Hickmann2002}. Polarization-sensitive
transmitter and receiver horns were used to couple microwaves between free
space and an HP 8720A vector network analyzer (VNA). The crystal was
positioned inside a microwave-shielded box with an aperture 14 cm $\times $
17 cm, deep in the far field (1.6 m) of the transmitter horn. The crystal is
oriented so that the microwaves are incident in the $\Gamma M$ direction 
\cite{Foteinopoulou2001}. The receiver horn was placed inside the box,
centered directly behind the crystal in line with the transmitter and
aperture. Since the crystal was far from the transmitter compared with the
relevant wavelengths, the microwaves incident upon the crystal were
effectively plane waves with a well-defined linear polarization. We confirmed
this by measuring the transmission between crossed horns in the absence of
the photonic crystal, finding a power suppression of $\geq $ 35 dB relative to
the aligned configuration. A schematic of the experimental setup is
presented in Fig. \ref{setup}.

\begin{figure}[h]
\centerline{\includegraphics{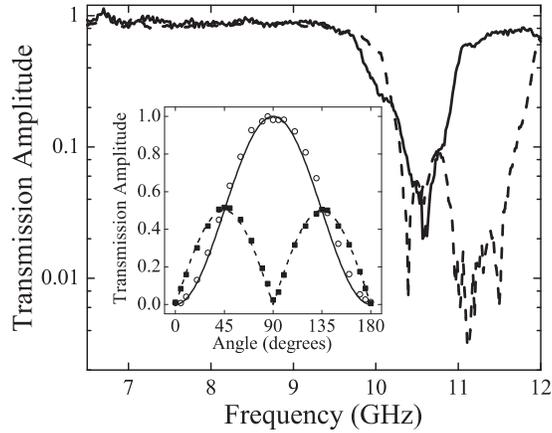}}
\caption{Experimentally measured electric field amplitude transmission spectrum for TE (solid
line) and TM (dashed line) waves. The inset shows field amplitude transmission data
(normalized to unity transmission) as a function of rotation angle in the
range 11.4-11.5 GHz for the horn antennae parallel (open circles) and
perpendicular (solid squares). Also plotted are the curves ${\sin }^{2}(%
\protect\theta )$ (solid line) and $\frac{1}{2}\left| {\sin }(2\,\protect%
\theta )\right|$ (dashed line).}
\label{LP-a}
\end{figure}

The fundamental band gap for our crystal is centered at roughly 11 GHz for
transverse magnetic (TM) waves and 10.5 GHz for transverse electric (TE)
waves. We label the polarizations TM and TE with respect to the plane of
periodicity of the crystal; thus the electric field is perpendicular to the
pipes for TM polarized waves and parallel to the pipes for TE polarized
waves. The band gap depth and width also depend on polarization. There is a clear
region of non-overlap around 11.4 GHz, in which TM radiation is strongly
reflected and TE radiation is primarily transmitted (see Fig. \ref{LP-a}).

In order to verify the operation of the crystal as a linear
polarizer, we fixed the crystal orientation and simultaneously rotated both horns around
the axis of propagation. If the horns are kept parallel during the rotation, this is equivalent 
to fixing the horns and rotating
the crystal. Since the horns are polarization-sensitive, this operation
should give a measured transmitted field of%
\begin{equation}
E=E_{0}~{\sin }^{2}~\theta \text{,}  \label{sinsqrd}
\end{equation}%
where $E_{0}$ is the incident electric field amplitude and  $\theta $ is the
rotation angle defined such that at $\theta =0$ the radiation detected by
the receiver horn corresponds to pure TM polarized waves. We also performed
this experiment with the horns perpendicular (crossed), in which case
simultaneous rotation gives%
\begin{equation}
E=E_{0}\left\vert \cos (\theta )\,\sin (\theta )\right\vert =\frac{1}{2}%
\left\vert {\sin }(2\,\theta )\right\vert \text{.}  \label{sincos}
\end{equation}

The results of these experiments, normalized to a maximum of unity
transmission, are shown in the inset of Fig. \ref{LP-a}. In both cases, the
data agree very well with the expected polarizer relations of Eqs. 1 and 2.
The intensity transmission at 0 degrees with the horns parallel was 0.0036\%
of the incident radiation, and the maximum to minimum transmitted power
ratio was roughly 15,000:1, limited by our experimental resolution.

We note that the maximum external transmission of TE polarized
radiation at 11.4 GHz was only $\approx 80\%$. This power loss is due in part to external 
reflections from the crystal faces. The remainder arises because the
transmission through a photonic crystal at fixed frequencies outside the
band gap displays oscillatory properties as a function of the number of
layers in the structure. These effects do not represent a fundamental limit
to the quality of photonic crystal linear polarizer devices, because with
proper engineering and a greater number of layers these values will improve.

\begin{figure}[h]
\centerline{\includegraphics{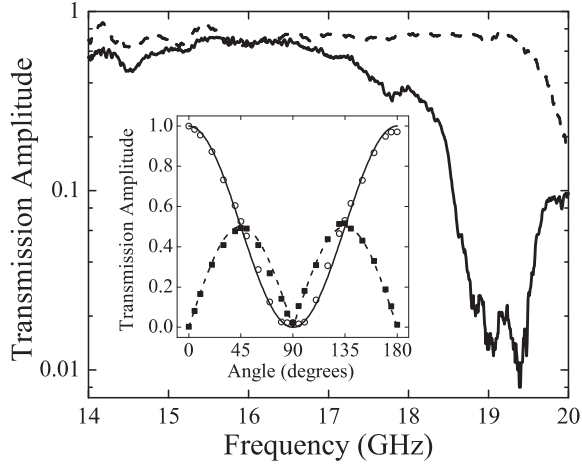}}
\caption{Experimentally measured electric field amplitude transmission spectrum for TE (solid
line) and TM (dashed line) waves. The inset shows field amplitude transmission data
(normalized to unity transmission) as a function of rotation angle in the
range 19.1-19.2 GHz for the horn antennae parallel (open circles) and
perpendicular (solid squares). Also plotted are the curves ${\cos }^{2}(%
\protect\theta )$ (solid line) and $\frac{1}{2}\left| {\sin }(2\,\protect%
\theta )\right|$ (dashed line).}
\label{LP-b}
\end{figure}

In addition to the fundamental band gap, we observed the next higher order
or double-frequency band gap in our structure (see Fig. \ref{LP-b}). This
gap is centered at roughly twice the frequency of the fundamental and has
approximately twice the width. Its average depth appears greater, which can
be qualitatively understood based on the increased optical path length at
this frequency.

We repeated the above experiments between 19.1 and 19.2 GHz (within this
double-frequency band gap) to verify the operation of the linear polarizer
at higher-order band gaps. An interesting feature in this region is that the
band gaps have an appreciable non-overlap region to the red (in addition
to the one to the blue) in which TM radiation is transmitted and TE
radiation is reflected. Because of this inverted reflection/transmission
relation, the maximum transmission with the horns aligned should occur at 0
and 180$^{\circ }$ (using the same angle convention as before) rather than 90%
$^{\circ }$. With the horns crossed, the angular dependence should again be
given by Eq. \ref{sincos}. The results of these measurements (again
normalized to a maximum of unity transmission) are shown in Fig. \ref{LP-b}
along with the relevant TM and TE transmission spectra. In both cases, the
data show the expected angular dependence. With the horns aligned, the
intensity transmission at 90 degrees was 0.032\% of the incident radiation,
and the maximum to minimum power ratio was approximately 1,800:1.

It is worth noting that the photonic crystal polarizer discussed here
differs from conventional linear polarizers because it reflects rather than
absorbs the rejected polarization. In this sense, it is similar to a PBS.
However, unlike conventional PBS devices, which operate by Brewster-angle
reflections, this beam splitter can be engineered to split the polarizations
in either order. As demonstrated above, by appropriately choosing a
frequency region (or engineering a photonic crystal given a
fixed frequency) the beam splitter can transmit either TE or TM radiation,
while reflecting the other. Since this beam splitter does not rely on
Brewster's angle, the angle between the direction of propagation of the
rejected polarization and the incident beam is not constrained. In fact, as demonstrated here,
it works even at normal incidence. Because of the previous considerations,
photonic crystal PBSs are
much more versatile devices than conventional, Brewster-based beam splitters.

In designing photonic crystal polarizers and PBSs, it will be important to
find or engineer structures with the correct band structure characteristics.
To be useful as a polarizer, a photonic crystal should have a deep and wide
band gap for at least one polarization. If the structure has band gaps for
both independent polarizations, there must exist some frequency region in
which they do not overlap, and the wider this region, the greater the useful
bandwidth of the polarizer. Along these lines, it is also advantageous to
utilize higher-order band gaps. They are generally deeper than the fundamental, and
since their center frequencies occur at multiples of the fundamental
frequency, their widths and relative separation for the two polarizations
are greater.

In conclusion, we have demonstrated that photonic crystals can be used as
linear polarizers and polarizing beam splitters by exploiting their
polarization-dependent transmission and reflection properties. Since
Maxwell's equations are scale-invariant, all the microwave results presented
here apply equally across the EM spectrum. We anticipate that these devices
will have applications in situations requiring compact, tailored
polarization control. They may be particularly useful in the optical region
of the spectrum as integrated elements in photonic crystal circuits and
related devices.

This work was supported by ARO grant number DAAD19-02-1-0276. We thank the
UC Berkeley Astronomy Department, in particular Dr. R. Plambeck, for lending
us the VNA. JMH thanks the support from Instituto do Mil\^{e}nio de Informa%
\c{c}\~{a}o Qu\^{a}ntica, CAPES, CNPq, FAPEAL, PRONEX-NEON, ANP-CTPETRO.

\end{document}